\documentstyle[12pt]{article}
\newcommand{\Ibb}[1]{ {\rm I\ifmmode\mkern
            -3.6mu\else\kern -.2em\fi#1}}
\newcommand{\ibb}[1]{\leavevmode\hbox{\kern.3em\vrule
     height 1.2ex depth -.3ex width .2pt\kern-.3em\rm#1}}

\newcommand{\Rl}{{\Ibb R}}
\newcommand{\be}{\begin{eqnarray}}
\newcommand{\ee}{\end{eqnarray}}

\newcommand{\A}{{\cal A}}

\renewcommand{\d}{\mbox{d}}

\begin{document}

\begin{tabbing}
\hspace*{12cm} \= GOET-TP 111/96  \\
               \> 1 August 1996
\end{tabbing}
\vskip1.cm

\centerline{\huge \bf Soliton Equations} 
\vskip.5cm
\centerline{\huge \bf  and the Zero Curvature Condition}
\vskip.5cm
\centerline{\huge \bf in Noncommutative Geometry}
\vskip1cm

\begin{center}
      {\large \bf Aristophanes Dimakis} \ and \ {\large \bf Folkert   
       M\"uller-Hoissen}
       \vskip.3cm 
      Institut f\"ur Theoretische Physik  \\
      Bunsenstr. 9, D-37073 G\"ottingen, Germany
\end{center} 
\vskip1cm

\begin{abstract}
\noindent
 Familiar nonlinear and in particular soliton equations arise as zero 
curvature conditions for $GL(1,\Rl)$ connections with noncommutative 
differential calculi. 
The Burgers equation is formulated in this way and the Cole-Hopf 
transformation for it attains the interpretation of a transformation 
of the connection to a pure gauge in this mathematical framework.
The KdV, modified KdV equation and the Miura transformation are 
obtained jointly in a similar setting and a rather straightforward 
generalization leads to the KP and a modified KP equation. 
 Furthermore, a differential calculus associated with the Boussinesq 
equation is derived from the KP calculus.
\end{abstract}
\vskip1cm

\section{Introduction}
Soliton equations are known to admit zero curvature 
formulations (see \cite{Fadd+Takh87}, for example). In case of the 
Korteweg-deVries (KdV), sine-Gordon and sinh-Gordon equations, one can 
find $SL(2,\Rl)$-connection 1-forms (gauge potentials) $A$ such that the 
condition of vanishing curvature (or `field strength')
\be                 		     \label{F-A}
             F := \d A + A A = 0  \; . 
\ee
is equivalent to the respective soliton equation \cite{Cra78}. 
In this work we show that the Burgers, KdV, Kadomtsev-Petviashvili
(KP) and Boussinesq equation can even be expressed as a zero curvature 
condition for a $GL(1,\Rl)$-connection, but with respect to a 
{\em noncommutative} differential calculus. 
By the latter we mean an analogue of the calculus of differential
forms on a manifold, but here functions and 1-forms in general do
not commute. As a consequence, the product of a 1-form with itself 
need not vanish, in contrast to the case of
the ordinary differential calculus. Because of this fact, one already
obtains nontrivial equations from $F = 0$ for a single 1-form $A$
(and not just for a matrix of 1-forms). 
\vskip.2cm

The relevant mathematical framework underlying this work is the theory
of differential calculi on commutative algebras. An exposition to it
can be found in \cite{BDMH95}, see also the references therein.
In `noncommutative geometry' an associative and not necessarily
commutative algebra replaces the algebra of (smooth) functions on
a manifold. A differential calculus on the algebra is then regarded
as the most basic geometric structure on which further geometric
concepts like connections can be defined. Though in this paper
we still deal with commutative algebras and thus topological spaces,
nontrivial commutation relations are introduced between functions 
and (generalized) 1-forms and this catches already much of the flair
of general noncommutative geometry.
\vskip.2cm

Irrespective of the choice of a differential calculus, the
expression (\ref{F-A}) makes sense. A gauge 
transformation with an invertible function $\psi$ is given by
\be	   \label{gauge-trafo}
   A' = \psi \, A \, \psi^{-1} - \d \psi \, \psi^{-1}   
\ee 
and induces the transformation $F' = \psi \, F \, \psi^{-1}$ of the
curvature 2-form.\footnote{More generally, this holds for any
$G$-connection where $G$ is a matrix group in which $\psi$ has values.}
\vskip.2cm

In section 2 the Burgers equation is obtained via a zero curvature 
condition with respect to a simple deformation of the ordinary 
differential calculus, and the Cole-Hopf transformation
appears as a transformation to a pure gauge. In section 3 we
start with two differential operators which play a role in the theory
of the Korteweg-deVries	(KdV) equation. We construct a differential
calculus in which these operators appear as generalized partial
derivatives. From the zero curvature condition for a $GL(1,\Rl)$
connection we then recover the KdV, modified KdV equation and the
Miura transformation. 
Even more interesting is the fact that our treatment of the KdV
equation naturally leads to the KP equation via a dimensional
continuation of the differential calculus associated with the
KdV equation. This is the subject of section 4.
In section 5 we show how the Boussinesq equation and its 
associated differential calculus arise via a dimensional reduction 
of the calculus associated with the KP equation. 
Section 6 contains some conclusions.

\section{The Burgers equation} 
In the following, $\A$ denotes the algebra of 
$C^\infty$-functions on $\Rl^2$. $t$ and $x$
are the canonical coordinate functions on $\Rl^2$.
Let $\Omega(\A)$ be the differential calculus determined by
\be
     \lbrack \d t , t \rbrack = \lbrack \d x , t \rbrack 
  = \lbrack \d t , x  \rbrack = 0 \, , \quad 
    \lbrack \d x , x \rbrack = \eta \, \d t      \, ,
\ee
with a constant $\eta$. More generally, we have
\be
      \d t \, f = f \, \d t  \, ,  \quad  
      \d x \, f = f \, \d x + \eta \, f_x \, \d t    
\ee
for $f \in \A$. Here $f_x$ denotes the partial derivative with
respect to $x$. For the differential of a function one obtains
\be 
  \d f = ( f_t + {\eta \over 2} \, f_{xx} ) \, \d t + f_x \, \d x
\ee
and 
\be
   \d x \, \d x = 0 = \d t \, \d t \, , \qquad
   \d x \, \d t = - \d t \, \d x  \; .
\ee 
This calculus has already been explored in several papers, see
\cite{DMH93} in particular. The differentials $\d t$ and $\d x$ 
constitute a basis of the space of 1-forms $\Omega^1(\A)$
as a left or right $\A$-module. 
\vskip.2cm

A $GL(1,\Rl)$-connection 1-form can be written as
\be
    A = w \, \d t + u \, \d x
\ee
with functions $u, w$. Using the differential calculus introduced
above, the curvature becomes
\be
    F = ( u_t + {\eta \over 2} \, u_{xx} + \eta \, u \, u_x  - w_x )
        \, \d t \, \d x  \; .
\ee
 For $w = 0$ the zero curvature condition takes the form 
\be
    u_t + {\eta \over 2} \, u_{xx} + {\eta \over 2} \, (u^2)_x = 0  
\ee
which is the Burgers equation \cite{Burg48,Draz+John89}.
On the other hand, it is easily verified that the zero curvature 
condition implies that $A$ can be written as a `pure gauge', i.e.
\be
  A = \theta^{-1} \, \d \theta = \theta^{-1} \, \left( \lbrack
      \theta_t + {\eta \over 2} \theta_{xx} \rbrack \, \d t + \theta_x
      \, \d x \right) 
\ee
with an invertible function $\theta$. Comparing the last expression
with $A = u \, \d x$, we obtain 
\be
  u = \theta_x / \theta  \; , \qquad \theta_t 
      + {\eta \over 2} \, \theta_{xx} = 0 \; .
\ee
Here we rediscover the Cole-Hopf transformation \cite{Draz+John89} 
which reduces the Burgers equation to the linear diffusion equation.

\section{The KdV equation}
A starting point in a modern treatment of the KdV equation is given
by the two `undressed' (cf \cite{Draz+John89}) differential operators
\be
    \Delta^{(0)}_1 := \partial_t + a b \, \partial_x^3 \; , \qquad
    \Delta^{(0)}_2 := b \, \partial_x^2
\ee
with nonvanishing real constants $a$ and $b$. 
Let us try to understand these as generalized partial derivatives
of a (noncommutative) differential calculus on $\A = C^\infty(\Rl^2)$. 
The associated exterior derivative should then act on a function 
$f \in \A$ as follows,
\be			     \label{KdV-d}
    \d f = (f_t + a b \, f_{xxx}) \, \d t
           + b \, f_{xx} \, \xi + f_x \, \d x
\ee
where $\xi$ is a 1-form which together with $\d t$ and $\d x$
constitutes a basis of the space of 1-forms as a left $\A$-module.
The operator $\d$ has to satisfy the Leibniz rule
\be
    \d (f h) = (\d f) \, h + f \, \d h  
\ee  
for $f,h \in \A$. It leads to
\be
  & & (f_t + a b \, f_{xxx}) \, \lbrack h , \d t \rbrack
   + b \, f_{xx} \, ( \lbrack h , \xi \rbrack + 3 a
   \, h_x \, \d t )  \nonumber \\
  & & + f_x \, ( \lbrack h , \d x \rbrack + 3 a b \, h_{xx} \, 
   \d t + 2 b \, h_x \, \xi ) = 0
\ee
This is only satisfied (for all smooth functions $f,h$) if 
the following commutation relations between functions and 1-forms
hold,
\be
      \d t \, f = f \, \d t  \, ,  \quad  
      \xi  \, f = f \, \xi + 3 a \, f_x \, \d t \, , \quad
      \d x \, f = f \, \d x  + 2 b \, f_x \, \xi 
                  + 3 a b \, f_{xx} \, \d t  \; .  
\ee
In particular,
\be          \label{3rd_dc}
    \lbrack \d t , t \rbrack = \lbrack \d x , t \rbrack 
 = \lbrack \d t , x \rbrack = \lbrack \xi , t \rbrack = 0
     \, , \quad
    \lbrack \d x , x \rbrack = 2 b \, \xi \, \quad 
     \lbrack \xi , x \rbrack = 3 a \, \d t  \; .
\ee
 Furthermore, we have the 2-form relations
\be		  \label{2form-KdV}
      \d t \, \d t = \d t \, \d x + \d x \, \d t 
   = \xi \, \xi = \xi \, \d t + \d t \, \xi = \xi \, \d x 
     + \d x \, \xi = 0  \, , \quad 
      \d \xi = - {1 \over b} \, \d x \, \d x  \; .
\ee  
They are obtained by acting with $\d$ on the equations 
(\ref{3rd_dc}) and by commuting $x$ through the resulting
relations using (\ref{3rd_dc}).
\vskip.2cm

A connection 1-form can be written as
\be			     \label{KdV-A}
      A = w \, \d t + u \, \xi + v \, \d x 
\ee
with functions $u, v, w$. The curvature of $A$ is 
\be
 F &=& \lbrack - b \, w_{xx} - 2 b \, v \, w_x 
       + u_t + a b \, u_{xxx}
       + 3 a \, u \, u_x + 3 a b \, v \, 
       u_{xx} \rbrack \, \d t \, \xi   \nonumber \\
   & & + \lbrack - w_x + v_t + a b \, v_{xxx} 
       + 3 a b \, v \, v_{xx}  
       + 3 a \, u \, v_x \rbrack \, \d t \, \d x 
       \nonumber \\
   & & + (- u + b \, v_x + b \, v^2 )_x  \, \xi \, \d x 
       + ( - {1 \over b} \, u + v_x + v^2 ) \, \d x \, \d x
       \; .       \label{F=0_KdV}
\ee  
If $\d \xi \neq 0$, the set of 2-forms $\d t \, \xi , \, 
\d t \, \d x , \, \d t \, \d x , \, \xi \, \d x , \, \d x \, \d x $ 
span the space $\Omega^2(\A)$ of 2-forms as a left (and right) 
$\A$-module and will be assumed to be a left $\A$-module 
basis.\footnote{We still have the freedom to modify the differential 
calculus at the level of 2-form relations, i.e., to modify the 
generalized wedge product by setting certain products of 
differentials, like $\d x \, \d x$, to zero. The corresponding terms 
in (\ref{F=0_KdV}) then simply drop out. In general it is more natural
to proceed without such extra conditions. There is some
motivation, however, to impose on the 1-form $\xi$ the condition 
$\d \xi = 0$, see the following section.}  
The zero curvature condition now becomes
\be         
   - {1 \over b} \, u + v_x + v^2 &=& 0  \label{Miura}  \\
   v_t + a b \, v_{xxx} + 3 a b \, v \, v_{xx}  
   + 3 a \, u \, v_x - w_x        &=& 0 \, , \label{0curv1} \\
   u_t + a b \, u_{xxx} + 3 a \, u \, u_x  
   - b \, w_{xx} + b \, v \, ( 3 a \, u_x  
   - 2 \, w )_x  &=& 0    \label{0curv2}
\ee 
where the first equation reminds us of the Miura transformation
\cite{Miura68,Draz+John89}.
The third equation obviously decouples from the others if we choose
\be
     w_x = {3 \over 2} \, a \, u_{xx}   \; .
\ee
However, taking (\ref{Miura}) into account, one finds a more general 
solution of the decoupling problem, namely
\be		\label{KdV-decoupl}
     w_x = {3 \over 2} \, a \, u_{xx} + c \, v_x 
\ee
with a constant $c$. (\ref{0curv2}) then becomes
\be
     u_t - c \, u_x + 3 a \, u \, u_x - {1 \over 2} \, a b
     \, u_{xxx} = 0
\ee
which for 
\be
     a=-2 \, , \qquad   b=1  \, , \qquad c = 0   \label{ab-choice}
\ee
is the KdV equation
as given in \cite{Draz+John89}, for example. We observe that the 
parameter $c$ simply reflects the effect of a special Galilean 
transformation.\footnote{For the Galilean transformation $x'=x+ct, 
\, t'=t$ we have $\partial_{t'} = \partial_t - c \partial_x, \, 
\partial_{x'} = \partial_x$.} 
To summarize, the zero curvature equation together with the 
restriction (\ref{KdV-decoupl}) on the connection $A$ leads to
the KdV equation. 
\vskip.2cm

With the help of (\ref{Miura}), the equation (\ref{0curv1}) is 
turned into
\be
   v_t - c \, v_x - {1 \over 2} a b \, v_{xxx} 
   + 3 a b \, v^2 \, v_x = 0
\ee
from which we recover what is known as a `modified KdV equation'
\cite{Draz+John89}. It is surprising that both, the KdV and the
mKdV equation appear jointly in our mathematical framework.
\vskip.2cm
  
In the above differential calculus it is consistent to impose the
additional condition that the 1-form $\xi$ is closed, i.e.
$\d \xi = 0$. The above formulae remain valid, except that now
\be
   \d x \, \d x = 0  \; .
\ee
The zero curvature condition is then slightly
less restrictive. It still leads to (\ref{0curv1}) and 
(\ref{0curv2}), but (\ref{Miura}) is replaced by the weaker
equation
\be
     {1 \over b} \, u = v_x + v^2 + \lambda
\ee
with a function $\lambda(t)$. For constant $\lambda$ we rediscover
what is sometimes refered to as the `Miura-Gardner transformation'
(see \cite{Wils81}, for example).
\vskip.2cm

 From the gauge transformation rule (\ref{gauge-trafo}) we obtain
\be
    \d \psi = \psi \, A - A' \, \psi 
\ee
where $\psi$ is an invertible (smooth) function. Using (\ref{KdV-d})
and (\ref{KdV-A}) this becomes
\be
   \psi_t + ab \, \psi_{xxx} &=& (w - w') \, \psi - 3a \, u' \, \psi_x
                           - 3ab \, v' \, \psi_{xx}	\\
   b \, \psi_{xx} &=& (u - u') \, \psi - 2b \, v' \, \psi_x   \\
   \psi_x &=& (v - v') \, \psi    \; .
\ee
If $\d \xi = 0$, a simple solution of the zero curvature condition and 
(\ref{KdV-decoupl}) is given by $A' = \lambda \, \xi$ with a constant 
$\lambda$. This determines a trivial solution of the KdV equation.
The above equations then take the form
\be
 \psi_t + ab \, \psi_{xxx} &=& w \, \psi - 3a \, \lambda \, \psi_x \\
 b \, \psi_{xx} &=& (u - \lambda) \, \psi   \label{KdV-ls1} \\
 \psi_x &=& v \, \psi    
\ee
and enforce that $A$ also has vanishing curvature. Restricting the
gauge transformation further in such a way that $A$ satisfies
(\ref{KdV-decoupl}) and thus determines a solution of the KdV
equation, and making use of the last of the above equations, the 
first equation becomes
\be			     \label{KdV-ls2}
   \psi_t + ab \, \psi_{xxx} = ({3a \over 2} \, u_x + f) \, \psi 
                         + (c - 3a \, \lambda) \, \psi_x	
\ee
where an arbitrary function $f(t)$ arose from integration of 
(\ref{KdV-decoupl}). Using the Sturm-Liouville equation 
(\ref{KdV-ls1}), (\ref{KdV-ls2}) can be written as
\be
   \psi_t = ({a \over 2} \, u_x + f) \, \psi 
         + (c - a \, u - 2 a \lambda) \, \psi_x  \; .	
\ee
 For $a=-2, \, c=0$ and special choices of $f$ this is the 
time-evolution equation for eigenfunctions of the Schr\"odinger 
operator associated with the KdV equation (cf \cite{Draz+John89}, 
p. 101).

\section{The KP equation}
In the differential calculus introduced in the previous section
it is tempting to replace the 1-form $\xi$ by the differential
$\d y$ of a third coordinate function $y$. For functions $f$ which 
do not depend on $y$ we recover the formulae of the previous
section. But the extension to functions of $t,x,y$ requires
nontrivial modifications. A consistent differential calculus 
on $\A = C^\infty(\Rl^3)$ is obtained by supplementing the 
relations (\ref{3rd_dc}) with\footnote{This is really the 
minimal extension of (\ref{3rd_dc}) obtained via $\xi \mapsto \d y$.
Note that $\lbrack \d x , y \rbrack = \lbrack \d y , x \rbrack$
using the Leibniz rule and $\lbrack x , y \rbrack = 0$.}
\be
    \lbrack \d t , y \rbrack = \lbrack \d y , y \rbrack = 0 \; ,
    \qquad
    \lbrack \d x , y \rbrack = 3 a \, \d t  \; .
\ee
Then
\be
      \d t \, f = f \, \d t  \, ,  \quad  
      \d y \, f = f \, \d y + 3 a \, f_x \, \d t \, , \quad
      \d x \, f = f \, \d x  + 2 b \, f_x \, \d y  
      + 3 a ( f_y + b \, f_{xx} ) \, \d t    
\ee
and
\be
     \d f = ( f_t + 3 a \, f_{xy} + a b \, f_{xxx} ) \, \d t
            + ( f_y + b \, f_{xx} ) \, \d y + f_x \, \d x  \; .
\ee
The set of 2-form relations (\ref{2form-KdV}) is extended by
\be
   \d t \, \d y + \d y \, \d t = \d x \, \d y + \d y \, \d x 
   = \d x \, \d x = 0 
\ee
(and modified via $\xi = \d y$, of course). Now $\d t \, \d x,
\, \d t \, \d y , \, \d y \, \d x$ is a basis of the space of
2-forms as a left $\A$-module.
\vskip.2cm

Any 1-form $A$ can be expressed as
\be
     A = w \, \d t + u \, \d y + v \, \d x
\ee
with function $u,v,w$. Regarded as a $GL(1,\Rl)$ connection 1-form,
the curvature is
\be
  F &=& \d A + A \, A  \nonumber \\
    &=& \lbrace -w_y - b \, w_{xx} + u_t + 3 a \, u_{xy} 
        + a b \, u_{xxx}     \nonumber \\
    & & + 3a \, u u_x - 2 b \, v w_x + 3a \, v (u_y + b \, u_{xx})
        \rbrace \, \d t \, \d y   \nonumber \\
    & & + \lbrace -w_x + v_t + 3 a \, v_{xy} + a b \, v_{xxx}
        + 3a \, u v_x + 3 a \, v (v_y + b \, v_{xx}) \rbrace 
        \, \d t \, \d x		\nonumber \\
    & &	+ \lbrace - u_x + v_y + b \, v_{xx} + 2 b \, v v_x 
        \rbrace \, \d y \, \d x   \; .
\ee
This vanishes iff
\be
     u_x &=& v_y + b \, (v_x + v^2)_x  \label{KP-F0-1}  \\
     w_x &=& v_t + 3 a \, v_{xy} + a b \, v_{xxx}
             + 3a \, u v_x + 3 a \, v (v_y + b \, v_{xx})  \\
     w_y + b \, w_{xx} &=& u_t + 3 a \, u_{xy} + a b \, u_{xxx}
        + 3a \, u u_x - v \, ( 2 b \, w_x - 3a (u_y + b \, u_{xx}) )
           \; .		   \label{KP-F0-3}
\ee
The next step parallels that of the KdV case treated in the previous
section. $v$ is obviously eliminated from the last equation by setting
\be				
  w_x = {3a \over 2b} \, u_y + {3a \over 2} \, u_{xx} \; .
\ee  
Motivated by the KdV example, we shall consider the more general 
expression
\be				 \label{KP-decoupl}
  w_x = {3a \over 2b} \, u_y + {3a \over 2} \, u_{xx} + c \, v_x 
\ee
where $c$ is an arbitrary constant. Taking (\ref{KP-F0-1}) into 
account, (\ref{KP-F0-3}) then reduces to
\be
  w_y = u_t - c \, (u_x - v_y) + {3a \over 2} \, u_{xy} 
        - {ab \over 2} \, u_{xxx} + 3a \, u u_x  \; . \label{KP-F0-3r}
\ee
Now there is an integrability condition. Comparing the results
obtained by differentiating (\ref{KP-decoupl}) with respect to $y$ 
and (\ref{KP-F0-3r}) with respect to $x$, we obtain
\be
   (u_t - c \, u_x - {ab \over 2} \, u_{xxx} + 3 a \, u u_x)_x 
   - {3a \over 2b} \, u_{yy} = 0 	   \label{KP}
\ee
which is the Kadomtsev-Petviashvili (KP) equation (for the choices
(\ref{ab-choice}) of the constants $a,b,c$, see \cite{Draz+John89}
for example). Again, via a Galilean transformation the constant $c$ 
can be eliminated. Though on the level of the zero curvature equations
our ansatz (\ref{KP-decoupl}) with $c \neq 0$ does not really decouple 
the variables because of the term $c \, v_y$, the latter does not enter
the integrability condition. 
\vskip.2cm

Let us now turn to the equation for $v$ which resulted from the
zero curvature condition. Taking (\ref{KP-decoupl}) into account,
we have
\be
  {3a \over 2b} \, u_y &=& v_t - c \, v_x + {3 a \over 2} \, v_{xy} 
  - {a b \over 2} \, v_{xxx} - 3 a b \, v_x^2 + 3 a \, v v_y
  + 3 a \, u v_x  \; .	      \label{KP-u-v}
\ee
Expressing $v$ as
\be
     v = q_x
\ee
with a function $q$, (\ref{KP-F0-1}) becomes
\be
   u_x &=& q_{xy} + b \, (q_{xx} + q_x^2)_x
\ee
and thus
\be
   u &=& q_y + b \, (q_{xx} + q_x^2) + f   \label{KP-u-q}
\ee
where $f$ is a function which does not depend on $x$, i.e.
$f(t,y)$. Now we can eliminate $u$ from (\ref{KP-u-v}) and obtain
\be
   (q_t - c \, q_x - {ab \over 2} \, q_{xxx} + ab \, q_x^3)_x
  + 3a \, (q_y + f) \, q_{xx} - {3a \over 2b} (q_{yy} + f_y) = 0 \; .
\ee 
Expressing $f$ as $f=h_y$ with a function $h(t,y)$, a field 
redefinition $q \mapsto q - h$ eliminates $f$ from the last
equation and we get
\be
   (q_t - c \, q_x - {ab \over 2} \, q_{xxx} + ab \, q_x^3)_x
  + 3a \, q_y \, q_{xx} - {3a \over 2b} q_{yy} = 0 \; .
\ee
This equation may be called a `modified KP equation' (mKP). 
We note that
\be
    \mbox{KP} = (\partial_y + b \, \partial_x^2 + 2b \, q_x \,
    \partial_x + 2b \, q_{xx}) \; \mbox{mKP}   \; .
\ee
Hence, given a solution $q$ of the mKP equation, then $u$ determined
by (\ref{KP-u-q}) is a solution of the KP equation.

\section{The Boussinesq equation}
Restricting the KP equation (\ref{KP}) to the hypersurface $t=0$
and renaming $y$ as $t$ afterwards, we arrive at the equation
\be
   u_{tt} + {2bc \over 3a} \, u_{xx} - b \, (u^2)_{xx} 
   + {b^2 \over 3} \, u_{xxxx} = 0     	   \label{Bouss}
\ee
which includes the Boussinesq equation (see \cite{Draz+John89},
for example).
The differential calculus associated with this equation is obtained
as a reduction of the calculus which led us to the KP equation in
the last section. First, we have to replace $\d t$ by some `abstract'
1-form $\xi$. Then, keeping our renaming $y \mapsto t$ in mind, the
commutation relations defining the differential calculus of the
previous section yield
\be
   \lbrack \d t , t \rbrack &=&  \lbrack \xi , t \rbrack
    = \lbrack \xi , x \rbrack = 0  \nonumber  \\
   \lbrack \d t , x \rbrack &=& \lbrack \d x , t \rbrack 
   = 3a \, \xi \, , \quad
   \lbrack \d x , x \rbrack = 2b \, \d t   \; .
\ee 
The differential of a function $f$ is now given by
\be
   \d f = (3a \, f_{tx} + ab \, f_{xxx}) \, \xi + (f_t + b \, f_{xx})
   \, \d t + f_x \, \d x  \; .
\ee  
Of course, we could have started with the differential calculus
determined by these relations and derived the Boussinesq equation
in the same way as we derived the KdV equation in section 3. 
In this case we in fact need to add a term proportional to $v_x$
in the decoupling ansatz for $w_x$ in order to recover the $u_{xx}$
part of (\ref{Bouss}). The term was of minor importance in our 
previous examples (see (\ref{KdV-decoupl}) and (\ref{KP-decoupl})).

\section{Conclusions}
Crucially underlying this work is the observation that with respect
to a noncommutative differential calculus already the field strength
(curvature) of a single connection 1-form (i.e., a $GL(1,\Rl)$ or a
$U(1)$ connection) involves nonlinear terms. With the ordinary
calculus of differential forms on a manifold, a matrix of connection
1-forms is needed to achieve that. This observation suggested to
investigate which well-known nonlinear field equations and 
in particular soliton equations can be formulated as zero curvature
conditions for a single connection 1-form with respect to a suitable
noncommutative differential calculus. We found that the
Burgers, KdV, KP and Boussinesq equation indeed admit such a
formulation. Of most interest is the fact that the differential
calculus associated with the KdV equation has a natural extension
and, following the steps which led to the KdV equation, we are led
straight to the KP equation.
\vskip.2cm

Nevertheless, though things fit surprisingly well together, a
deeper understanding why this is so is still lacking. 
 For a certain class of completely integrable models noncommutative
differential calculus has indeed led to a rather complete 
understanding and a receipe to construct new integrable models
\cite{DMH95,DMH96}. There is therefore much hope to achieve a 
comparable understanding of the structures presented in this work.

\end{document}